\documentclass[aps,prl,twocolumn]{revtex4}
\usepackage{bm,color,amsmath,amssymb,mathrsfs,latexsym,graphicx,psfrag,appendix}

\newcommand{\up}{\uparrow}
\newcommand{\dw}{\downarrow}

\newcommand{\bk}{{\mathbf k}}
\newcommand{\bq}{{\mathbf q}}
\newcommand{\br}{{\mathbf{r}}}

\def\be{\begin{equation}}
\def\ee{\end{equation}}
\def\bea{\begin{eqnarray}}
\def\eea{\end{eqnarray}}
\def\bD{{\mathbf{D}}}
\def\tr{\textrm{tr}}

\begin{document}

\title{New class of topological superconductors protected by magnetic group symmetries}
\author{Chen Fang$^{1,3,4}$, Matthew J. Gilbert$^{2,3}$, B. Andrei Bernevig$^4$}
\date{\today}
\affiliation{$^{1}$Department of Physics, University of Illinois, Urbana IL 61801-3080}
\affiliation{$^2$Department of Electrical and Computer Engineering, University of Illinois, Urbana IL 61801}
\affiliation{$^3$Micro and Nanotechnology Laboratory, University of Illinois, Urbana IL 61801}
\affiliation{$^4$Department of Physics, Princeton University, Princeton NJ 08544}
\begin{abstract}
We study a new type of three-dimensional topological superconductors that exhibit Majorana zero modes (MZM) protected by a magnetic group symmetry, a combined antiunitary symmetry composed of a mirror reflection and time-reversal. This new symmetry enhances the noninteracting topological classification of a superconducting vortex from $Z_2$ to $Z$, indicating that multiple MZMs can coexist at the end of one magnetic vortex of unit flux. Specially, we show that a vortex binding two MZMs can be realized on the $(001)$-surface of a topological crystalline insulator SnTe with proximity induced BCS Cooper pairing, or in bulk superconductor In$_x$Sn$_{1-x}$Te.
\end{abstract}
\maketitle

Topological superconductors are superconductors that are fully gapped in the bulk and yet possess gapless boundary excitations at zero-dimensional\cite{Read2000,Kitaev2001,Fu:2008fk} (0D), one-dimensional (1D) or two-dimensional\cite{schnyder2008,Kitaev2009,Qi2010,Fu2010,Sato2010} (2D) boundaries, called Majorana zero modes (MZM), respectively\cite{Moore2010,qi2011rev,Bernevig2013}. 0D MZMs have so far received most attention\cite{Kitaev2001,Sato2009,Sau2010,Lutchyn2010,Moore2010,Alicea2011,Mourik2012,Das2012}. The proposals of realizing these states include vortex bound states in $p+ip$ superconductors\cite{Read2000}, vortex bound states on surfaces of a strong topological insulator (TI) with induced $s$-wave superconductivity\cite{Fu:2008fk} and the end states of a spin-orbital coupled quantum wire with proximity induced $s$-wave superconductivity and subject to a strong Zeeman field\cite{Sato2009,Sau2010,Lutchyn2010,Alicea2011,Mourik2012}.

In above proposals of 0D MZM, a \emph{single} MZM exists, while any two Majorana modes will hybridize and open a gap. In the presence of two or more Majorana modes, perturbations in the form $\Delta{H}=i\gamma_a\gamma_b$ can be added, where $\gamma_{a}$ denotes a Majorana fermion operator of species $a$. Such perturbations gap Majorana modes in pairs, giving $Z_2$ topological classification of 0D systems with no symmetry. Existence of multiple Majorana modes requires symmetry to forbid their hybridization. For example, with (spinful) time-reversal symmetry (TRS), a pair of MZMs can appear if the they make a Kramers' pair\cite{schnyder2008,Kitaev2009,Qi2010}. Local and unitary symmetries in general enhance the classification to $Z^k$, where $2k$ is the number of \emph{complex} eigenvalues of the symmetry operator\cite{Ueno2013,Chiu2013,Zhang2013,Morimoto2013}. The nontrivial phases in these classes require intrinsic or induced unconventional superconductivity, with sign changes in the pairing amplitude between different Fermi surfaces.

Here we propose a new class of 3D topological superconductors that have multiple MZMs bound to each magnetic vortex core of unit flux on certain surface terminations. The hybridization between MZMs is prohibited by a nonlocal magnetic group symmetry: a vertical mirror plane reflection followed by TRS, denoted by $M_T$. This is the generic symmetry of any superconductor that (i) has a mirror symmetric lattice, (ii) has mirror symmetric and TRS invariant Cooper pairing and (iii) is subject to an external magnetic field and/or Zeeman field parallel to the mirror plane. Neither mirror reflection nor TRS is a symmetry as they both invert the magnetic/Zeeman field which is a pseudo-vector, but their combination leaves the field invariant. This symmetry was first identified by Tewari and Sau\cite{Tewari2012} as a `new TRS' in quasi-1D superconducting quantum wires with Zeeman field along the length, which can protect multiple MZMs, but is absent when inter sub-band Rashba coupling is included; in Ref.[\onlinecite{Mizushima2013}], a spin-orbital coupled quasi-1D optical lattice is proposed which has this exact symmetry and therefore hosts multiple MZMs. In this Letter, we prove that the topological classification protected by $M_T$ is $Z$ in general, and then we show that a $z=2$ state (having two protected MZMs at each vortex core) can be realized on the $(001)$-plane of topological crystalline insulator\cite{Fu2011,Hsieh2012,Fang2012} (TCI) SnTe with induced or intrinsic $s$-wave superconductivity on the surface. We expect that this phase can be realized in a $(001)$-thin-film SnTe deposited on an BCS-superconducting substrate such as NbSe$_2$ or bulk superconductor In$_x$Sn$_{1-x}$Te.

In the type-II limit, the magnetic field penetrates into the superconductor in the form of vortex lines along the field direction. We take the limit where vortex lines are far away from each other and can be considered isolated. Now we terminate the system on a surface perpendicular to the mirror plane. A terminated vortex line has the particle-hole symmetry (PHS) and the magnetic group symmetry $M_T$. Assume that the end of a vortex line hosts several MZMs close to each other. Their Hamiltonian can be written in the PHS symmetric basis (Majorana basis) as
\bea
\hat{H}=i\sum_{a,b}\mathcal{H}_{ab}\gamma_a\gamma_b,
\eea
where $\mathcal{H}_{ab}$ is a real skew-symmetric matrix. A matrix representation of $M_T$ is in general
\bea
M_T=K\mathcal{M},
\eea
where $\mathcal{M}$ is a unitary matrix and $K$ is complex conjugation. Physically, we have the following constraints on the form of $M_T$: (i) it must commute with PHS and (ii) $M_T^2=M^2\times{T}^2=-1\times(-1)=1$, as both mirror reflection and time-reversal square to $-1$ for a spinful fermion. (Here $M$ and $T$ represent operators for mirror reflection and TRS, respectively, in the single fermion Hilbert space.) They require that $\mathcal{M}$ be real and symmetric. Hence, the eigenvalues of $\mathcal{M}$ can only be $\pm1$. If $M_T$ is a symmetry of $\hat{H}$, we have
\bea\label{eq:general_constraint}
[i\mathcal{H},K\mathcal{M}]=\{\mathcal{H},\mathcal{M}\}=0.
\eea
Eq.(\ref{eq:general_constraint}), after straightforward algebraic work\cite{SupMat}, leads to the result that there are exactly $|\tr(\mathcal{M})|$ eigenvalues of $\mathcal{H}$ fixed at zero. Since $|\tr(\mathcal{M})|$ is an integer, there can be an integer number of MZMs at each end of the vortex line, giving rise to a $Z$-classification.

\begin{figure}
\includegraphics[width=8cm]{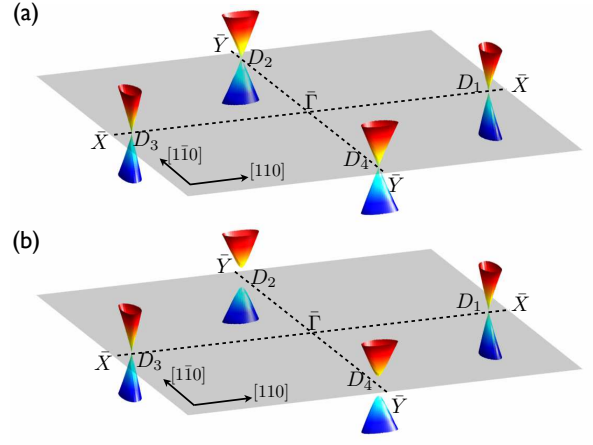}
\caption{(a) The dispersion of rock-salt SnTe $(001)$-surface bands, calculated using the model Eq.(\ref{eq:TB}) with $t_1=-1,\;t_2=0.5,\;m=2.5$. (b) The dispersion of SnTe $(001)$-surface bands with rhombohedral distortion along $[111]$-direction, the strength of which is $\epsilon=0.1$.}
\label{fig:SBZ}
\end{figure}

We have yet to determine the physical requirements, including band structure and the form of Cooper pairing, for a nontrivial superconductor that support such vortices to appear. In this Letter, we do not provide a general answer to this question, but instead provide a realization for each nontrivial phase in a class of hetero-structures made of conventional BCS-superconductors and new materials called topological crystalline insulators (TCI) having mirror symmetry and TRS. In TCI, the surface states have multiple Dirac points protected by mirror symmetries, if the surface termination is perpendicular to the mirror plane. Rock-salt (Pb,Sn)Te is a TCI having four Dirac cones on the $(001)$-surface\cite{Xu2012,Tanaka2012,Dziawa2012}: two along $k_y=0$ (denoted by $\bD_{1,3}$) and two others along $k_x=0$ ($\bD_{2,4}$), protected by mirror planes of $M_{1\bar10}$ and $M_{110}$, respectively. At the vicinity of each Dirac point, the low energy effective theory is that of 2D massless Dirac fermions [Fig.\ref{fig:SBZ}(a)]. We then assume a Cooper pairing induced on the surface states as preserves all lattice symmetries and TRS. In reality, this surface superconductivity can be proximity-induced by a conventional BCS superconductor. We prove that given the induced superconductivity in SnTe, any vortex line along $\{001\}$ has two MZMs protected by $M_T$. Fourfold symmetry that is specific to this system can be broken without changing the result. An extension of the discussion\cite{SupMat} applies to a general TCI with induced superconductivity, showing that there are exactly $|C_M|$ MZMs at the end of a vortex line protected by $M_T$, where $C_M$ is the mirror Chern number of the TCI.

First we consider the surface states in the normal state. At Dirac point $\bD_1$, the effective Hamiltonian is in general given by $\hat{h}_1=\sum_{|\bq|<\Lambda,\tau,\tau'=\up,\dw}h_1^{\tau\tau'}(\bq)f^\dag_{1\tau}(\bq)f_{1\tau'}(\bq)$, where $f_{1\tau}(\bq)$ is the annihilation operator at $\bk=\bD_1+\bq$ with pseudo-spin $\tau$, denoting each state of the degenerate doublet at $\bD_1$. The form of $h_1(\bq)$ is fixed by choosing the representation of the little group at $\bD_1$\cite{Fang2013} to be $M_{1\bar10}=i\sigma_y$ and $C_{2T}=K\sigma_x$, where $C_{2T}=C_2*T$ a twofold rotation followed by TRS. Using the symmetry constraint $[C_{2T},\hat{h}_1]=[M_{1\bar10},\hat{h}_1]=0$, we have:
\bea\label{eq:h1}
h_1(\bq)=v_0q_x\sigma_0+v_1q_x\sigma_y+v_2q_y\sigma_x
\eea
up to the first order of $|\bq|$. Here the sign of $v_1$ is determined by the sign of $C_M$, while other parameters are related to details of the system. Using $C_4$ symmetry, we can fix the gauge for Dirac cones centered at $D_{2,3,4}$: $f_{2,3,4}(C_4\bq)\equiv{}C_4f_{1,2,3}(\bq)C_4^{-1}$, where $C_4\bq$ is $\bq$ rotated by $\pi/2$, by which we have
\bea\label{eq:h234}
h_1(\bq)=h_2(C_4\bq)=h_3(C_2\bq)=h_4(C_4^{-1}\bq).
\eea
In Table \ref{tab:transform}, we list how $f_{i\tau}$ transforms under $C_{4v}\otimes{T}$, the full symmetry group of the $(001)$-plane.

\begin{table}[htdp]
\begin{center}
\begin{tabular}{|c|c|c|c|c|}
\hline
 & $f_1$ & $f_2$ & $f_3$ & $f_4$\\
\hline
$M_{1\bar10}$ & $(i\sigma_y)f_1$ & $(-i\sigma_y)f_4$ & $(-i\sigma_y)f_3$ & $(-i\sigma_y)f_2$\\
\hline
$M_{110}$ & $(-i\sigma_y)f_3$ & $(i\sigma_y)f_2$ & $(-i\sigma_y)f_1$ & $(-i\sigma_y)f_4$\\
\hline
$T$ & $-\sigma_x{f}_3$ & $-\sigma_x{f}_4$ & $\sigma_x{f}_1$ & $\sigma_x{f}_2$\\
\hline
$C_4$ & $f_2$ & $f_3$ & $f_4$ & $-f_1$\\
\hline
\end{tabular}
\end{center}
\caption{Transformation of operators under $C_{4v}$ and TRS.}
\label{tab:transform}
\end{table}%

Next we consider the Cooper pairing on the surface that does not break any lattice symmetry or TRS, such as that induced by a conventional BCS superconductor. A generic expression of a Cooper pairing with zero momentum is $\hat{\Delta}=\sum_{\bq}f_1^T(\bq)\Delta_Xf_3(-\bq)+f_2^T(\bq)\Delta_Yf_4(-\bq)+h.c.+O(|\bq|)$. Here we note that, since $f_{1,2,3,4}(\bq)$ carry momentum around $\bD_{1,2,3,4}$, other inter-cone pairings and intra-cone pairings are not allowed as both lead to paris of finite total momentum. Using Table \ref{tab:transform}, we find the only possible form of $\Delta_{X,Y}$ that preserves all symmetries is
\bea\label{eq:DeltaXY}
\Delta_X=\Delta_Y=\Delta_0\sigma_x,
\eea
where $\Delta_0$ is a real number representing the pairing amplitude. Combining Eq.(\ref{eq:h1},\ref{eq:h234},\ref{eq:DeltaXY}), we obtain the BdG Hamiltonian as
\bea\label{eq:fullH}
\hat{H}_0&=&\sum_{|\bq|<\Lambda}\{\sum_{i}f_i^\dag(\bq)h_i(\bq)f_i(\bq)\\
\nonumber
&+&\Delta_0[f_1^T(\bq)\sigma_xf_3(-\bq)+f_2^T(\bq)\sigma_xf_4(-\bq)+h.c.]\}\\
\nonumber&=&\sum_{\br}\{\sum_{i}f_i^\dag(\br)h_i(-i\nabla)f_i(\br)\\
\nonumber
&+&\Delta_0[f_1^T(\br)\sigma_xf_3(-\br)+f_2^T(\br)\sigma_xf_4(-\br)+h.c.]\},
\eea
where in the second line we have defined $f_i(\br)\equiv\frac{1}{\sqrt{N}}\sum_{|\bq|<\lambda}f_i(\bq)e^{i\bq\cdot\br}$ in real space. Since $f_i(\bq)$ has momentum $\bD_i+\bq$, $f_i(\br)$ represents the slowly oscillating part of the wavefunction. The energy dispersion of Eq.(\ref{eq:fullH}) is
\bea\label{eq:dispersion}
E(\bq)=\pm\sqrt{(v_0q_y-\mu\pm\sqrt{v_1^2q_x^2+v_2^2q_y^2})^2+\Delta_0^2},
\eea
where each band fourfold degenerate. Eq.(\ref{eq:dispersion}) shows that the bulk is a fully gapped superconducting state for any parameter set with $\Delta_0\neq0$. A superconducting vortex is created by replacing the constant pairing amplitude $\Delta_0$ with a spatially varying function having winding number $+1$ (the case of $-1$ can be similarly discussed). Here we take
\bea\label{eq:substitute}
\Delta_0\rightarrow\Delta(r)e^{i\theta}
\eea
written in polar coordinates, where $\Delta(r)$ is a monotonic real function of $r$ that satisfies $\Delta(0)=0$ and $\Delta(\infty)=\Delta_0$. The vortex bound state(s) can be found by diagonalizing Eq.(\ref{eq:fullH}) after the substitution of Eq.(\ref{eq:substitute}). As we have mentioned, all parameters except the sign of $v_1$ can be adiabatically changed, so here we take $v_0=\mu=0$ and $-v_2=v_1\equiv{v}$ without changing the topological class of the vortex. For these parameters, the bound state problem can be solved analytically. 
There are four MZMs, given by
\bea\label{eq:explicit}
\gamma_1&=&\sum_\br(f_{1\dw}(\br)\pm{f}_{3\dw}(\br)+h.c.)e^{-\int_0^r|\Delta(r')|dr'},\\
\nonumber
\gamma_2&=&\sum_\br(e^{i\pi/4}f_{2\dw}(\br)\pm{}e^{i\pi/4}{f}_{4\dw}(\br)+h.c.)e^{-\int_0^r|\Delta(r')|dr'},\\
\nonumber
\gamma_3&=&\sum_\br(if_{3\dw}(\br)\mp{i}{f}_{1\dw}(\br)+h.c.)e^{-\int_0^r|\Delta(r')|dr'},\\
\nonumber
\gamma_4&=&\sum_\br(e^{i3\pi/4}f_{4\dw}(\br)\mp{e}^{i3\pi/4}{f}_{2\dw}(\br)+h.c.)e^{-\int_0^r|\Delta(r')|dr'},
\eea
where the upper/lower sign is taken if $\textrm{sign}[v_1\Delta_0]=+/-$ and the normalization factors are omitted. $M_T\equiv{M}_{xz}\times{T}$ is a symmetry of the system with vortex, and using Table \ref{tab:transform}, we obtain the matrix representation of $M_T=K\mathcal{M}$ in the basis furnished by $\gamma_{1,2,3,4}$:
\bea\label{eq:MT}
\mathcal{M}=\textrm{sign}[v_1\Delta_0]\left(\begin{matrix}
1 & 0 & 0 & 0\\
0 & 0 & 0 & 1\\
0 & 0 & 1 & 0\\
0 & 1 & 0 & 0
\end{matrix}\right).
\eea
Since $\tr(\mathcal{M}_{T})=2\textrm{sign}[v_1\Delta_0]$, there are two MZMs that are topologically protected by $M_T$. Although $\mathcal{M}_{T}$ is evaluated using the explicit forms of bound state solutions, its trace is a good quantum number invariant under any adiabatic change of parameters. While we have four MZMs given in Eq.(\ref{eq:explicit}), only two are protected. This is because one can write down a perturbation: $\Delta\hat{H}=i\lambda(\gamma_1\gamma_2+\gamma_2\gamma_3+\gamma_3\gamma_4+\gamma_4\gamma_1)$ that preserves $M_{T}$ and gaps two out of the four MZMs. This perturbation does not break the fourfold rotation symmetry either, which means that $C_4$ symmetry here does not lead to additional degeneracy. A more detailed study in Ref.[\onlinecite{SupMat}] shows that when the size of the vortex is far greater than the lattice constant, the above perturbation is very small and the other two modes will hence be very close to the zero energy.

We have so far assumed that the Fermi level is inside the bulk gap of the 3D system, making it sufficient to consider only the surface electrons. This assumption is impractical for experimentally realizing and measuring the MZMs because when the Fermi level is inside the bulk gap, the proximity-induced SC decays exponentially fast away from the interface, leaving too small a superconducting gap on the open surface for measurements. Therefore, we must find out if the above results hold when the Fermi level is inside the conduction/valence bands. Generically, a vortex line undergoes a quantum phase transition at some critical chemical potential $\mu_c$ inside the bulk bands, at which the MZMs localized at the two ends extend into the bulk of the line and hybridize\cite{Hosur2011,Chiu2011,Hung2013}. Below we numerically confirm this picture in 3D SnTe.

\begin{figure}
\includegraphics[width=8cm]{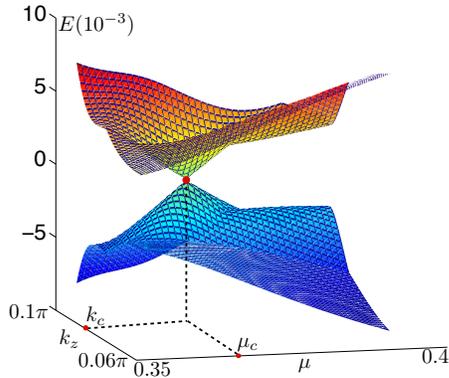}
\caption{The dispersion of the lowest bands in the vortex line as a function of $k_z$ and chemical potential $\mu$ close to the critical point where the band gap closes at $(k_c,\mu_c)$. Due to PHS, there is another band crossing at $(-k_c,\mu_c)$.}
\label{fig:QPT}
\end{figure}

We develop a 3D tight-binding model to describe the normal state of the TCI
\bea\label{eq:TB}
H(\bk)&=&[m-t_1(\cos{2k_x}+\cos{2k_y}+\cos{2k_z})]\Sigma_{z0}\\
\nonumber
&+&t_2[\sin{k_x}(\cos{k_y}+\cos{k_z})\Sigma_{xx}\\
\nonumber
&+&\sin{k_y}(\cos{k_x}+\cos{k_z})\Sigma_{xy}+\sin{k_z}(\cos{k_x}+\cos{k_y})\Sigma_{xz}],
\eea
where $\Sigma_{ij}\equiv\sigma_i\otimes\sigma_j$ and the parameters are chosen as $\{m,t_1,t_2\}=\{2.5,-1.0,0.5\}$. The model has cubic symmetry and TRS, and the BZ of this model is that of an FCC lattice (same as the real material); the symmetry group generators are given by the following matrices: $C_{4z}=\sigma_0\otimes{e}^{i\sigma_z\pi/4}$, $C_{4x}=\sigma_0\otimes{e}^{i\sigma_x\pi/4}$, $P=\Sigma_{z0}$ and $T=K(i\Sigma_{0y})$. The model gives the correct topological surface states as shown in Fig.\ref{fig:SBZ}. An onsite $s$-wave pairing with a vortex line is given by $\Delta(\br)=\Delta(\sqrt{x^2+y^2})e^{i\theta}(i\Sigma_{02})$. We take the simplest form of $\Delta(r)$: $\Delta(r)=0$ for $r<r_0$ and $\Delta(r)=\Delta_0$ for $r\ge{r_0}$. We solve the eigenvalue problem of a vortex line with periodic boundary along $z$-axis, and plot the energy spectrum against increasing chemical potential in Fig.\ref{fig:QPT}. The result shows that the phase transition happens at critical chemical potential $\mu_c>\mu_b$, where $\mu_b$ is the minimum of the conduction band. (In the particular parameter set we choose to calculate Fig.\ref{fig:QPT}, $\mu_c\approx0.38$ and $\mu_b\approx0.23$.) We note that there are two gap closings at $\mu_c$ with $k_z=\pm{k_c}$, in contrast to just one closing in Ref.[\onlinecite{Hosur2011}], because here the transition is between a vortex line having two MZMs and one having none.

Now we discuss the effect of the spontaneous rhombohedral distortion of SnTe at low temperatures, which has attracted theoretical and experimental attention\cite{Hsieh2012,Okada2013,Fang2013}. The lattice distortion is equivalent to a small strain tensor $\epsilon_{xy}=\epsilon_{yz}=\epsilon_{xz}=\epsilon$, which breaks both $C_2$ and $M_{110}$. It opens gaps at two Dirac points along $\bar\Gamma\bar{Y}$-direction, leaving the other two gapless, as $M_{1\bar10}$ is preserved. The strain gaps at $D_{2,4}$ can also be observed in our TB-model adding a perturbation (one could verify that it transforms the same way as the strain tensor under the point group) $\epsilon[(\cos2k_x-\cos2k_y)\sin{2k_z}+(\cos2k_y-\cos2k_z)\sin{2k_x}+(\cos2k_z-\cos2k_x)\sin{2k_y}]\Sigma_{y0}$ [see Fig.\ref{fig:SBZ}(b)]. However, this effect does not entail any topological transition in the vortex line. This is because the two MZMs at each end are protected by $M_T\equiv{M_{1\bar10}}\times{T}$, unbroken by the strain.

Based on the theory, we design a simple TCI-SC heterostructure to realize the nontrivial state with $z=2$. A thin-film SnTe is deposited on the top of a conventional superconductor such as NbSe$_2$. The Fermi level in the thin film is tuned through gating to a value inside but near the edge of the conduction/valence band ($\mu_b<|\mu|<\mu_c$). When the Fermi level is in the bulk bands, the proximity-induced SC pairings on the bottom layer of SnTe extend to the bulk with a power law decay. Therefore, on the top layer the SC pairing is still finite, and the whole thin film has an induced pairing that preserves all lattice symmetries and TRS. According to our theory, an isolated magnetic vortex in the thin film can bind exactly two MZMs on the top surface, which may be observed through tunneling measurements. Following a discussion similar to that presented in Ref.[\onlinecite{Roy2013}], we expect a zero bias conductance peak of intensity $4e^2/h$, if the tip is correctly located at the vortex. While the intrinsic rhombohedral strain as discussed above cannot open a gap between the MZMs, an applied strain which also breaks $M_{1\bar10}$ can break the double degeneracy, making the vortex line fully gapped, and the peak splits into two at nonzero voltages with intensity $2e^2/h$ each. The vortex bound MZMs may also be realized in superconducting In$_x$Sn$_{1-x}$Te, if the bulk superconducting gap is \emph{trivial}, in contrast to some theoretical proposals. The proposed nontrivial odd parity pairing in the bulk leads to 2D Majorana modes, while here we have shown that even if the bulk gap is trivial, two MZMs can still be observed at vortex lines that are parallel to the mirror planes of the crystal, given that the Fermi energy is in the inverted regime at the edge of the bulk bands.

For the general case of a non-interacting TCI with mirror Chern number $C_m$, we can similarly prove that a vortex line parallel to the mirror plane can bind exactly $C_m$ MZMs at each end. If an interaction, i.e., a four-Majorana term, be added to the system, the $Z$-classification of a vortex line reduces to $Z_8$ without breaking any symmetry. If $C_m=\pm4$, the noninteracting ground state is fourfold degenerate, but a four-Majorana interaction, in the form $\lambda\gamma_1\gamma_2\gamma_3\gamma_4$ lifts the degeneracy down to twofold. If $C_m=\pm8$, the noninteracting ground state is sixteen-fold degenerate, but a four-Majorana interaction that breaks the SO(8) symmetry (rotation symmetry in the flavor space) renders the many-body ground state non-degenerate\cite{Fidkowski2011}.

Finally, we discuss limitations of the theory. It presumes mirror symmetry in zero field, which is equivalent to the pure limit, because exact mirror symmetry is broken by any type of impurities. However, randomly distributed impurities may preserve mirror symmetry 'on average'\cite{Fu2012}. For this reason, the theory also applies in the case of many impurities if the size of the vortex is much larger than the average spacing of impurities, as long as the impurity intensity is much weaker than the superconducting gap. We also require the vortices to be sufficiently separated from each other such that the size of the Majorana fermions is much smaller than their average spacing. In our discussion, the magnetic field only supplies the vortex while the Zeeman field is ignored. The Zeeman field preserves $M_T$ and hence does not hybridize the MZMs, but we require that its strength not exceed the superconducting gap.

\textbf{Acknowledgements} CF thanks S. Xu, J. Maciejko and C.-X. Liu for helpful discussions. CF is supported by ONR-N00014-11-1-0728; CF acknowledges travel support from ONR-N00014-11-1-0635, and Darpa-N66001-11-1-411. MJG acknowledges support from the AFOSR under grant FA9550-10-1-0459 and the ONR under grant N0014-11-1-0728. BAB was supported by NSF CAREER DMR- 095242, ONR-N00014-11-1-0635, Darpa-N66001-11-1-4110, David and Lucile Packard Foundation, and MURI-130-6082.

\onecolumngrid
\appendix
\section{Proof that $|\tr(\mathcal{M})|$ equals the number of MZM's}
Using $\{\mathcal{H},\mathcal{M}\}=0$, we know that for any given eigenstate $|u\rangle$ of $M$ with eigenvalue $\pm1$, $\mathcal{H}|u\rangle$, if not a null vector, must be an eigenstate of $M$ with the eigenvalue $\mp1$. Denote by $N_\pm$ the number of eigenvalues of $M$ that have eigenvalues $+1$ and $-1$ respectively, and assume that $N_+>N_-$. The corresponding eigenstates are denoted by $|u_i^\pm\rangle$, forming subspaces $\Psi_\pm$. Obviously, since any nonzero $H|u_i^+\rangle$ for $i=1,...,N_+$ is a state in $\Psi_-$, there must be at least $N_+-N_-$ of them that are null vectors, i.e., $N_+-N_-$ eigenstates of $\mathcal{H}$ of zero eigenvalue. The same argument proceeds for the case $N_->N_+$, leading to the result that there must be at least $N_--N_+$ eigenstates of $\mathcal{H}$ of zero eigenvalue.

\section{MZM's in a general TCI having mirror Chern number $C_m$ with induced superconductivity}

In the main text, we explicitly show that on the $(001)$-surface of mirror TCI SnTe, there are two MZM's protected by $M_T$. Here we extend to all TCI's having (i) mirror Chern number $C_m$, (ii) TRS, (iii) Cooper pairing that preserves the above two symmetries and (iv) same sign of pairing amplitude on all pieces of the Fermi surface. All these make the system have both $M$ and $T$, so that when there is vortex, their combination $M_T$ is preserved, while they are no longer symmetries separately.

Given a mirror plane, a 3D BZ may have one (e.g., FCC) or two (e.g., simple cubic) mirror symmetric subspaces. For a surface termination that preserves the mirror symmetry, i.e., perpendicular to the mirror plane(s), there are one or two mirror symmetric lines in its SBZ. On a mirror symmetric line, bands with opposite mirror eigenvalues ($\pm{i}$ in spinful systems) can cross each other, leaving a set of Dirac points in the SBZ. On a mirror symmetric line in SBZ, TRS sends a right going mode to a left going mode while changing the mirror eigenvalue, making the number of right/left going modes with eigenvalue $+i$ exactly equals the number of left/right going modes with eigenvalue $-i$.

We now make an assumption that there is only one, instead of two, mirror symmetric plane in 3D BZ, or one mirror symmetric line in the SBZ. We will relax the assumption at a later stage. The mirror Chern number is defined as $C_m=n^+_R-n^+_L+n^-_L-n^-_R$, where $n^{+/-}_{L/R}$ is the number of left/right going modes with eigenvalue $+i/-i$. We can adiabatically tune the system such that (i) if $C_m>0$, all right going modes have $M=+i$ and all left going ones have $-i$ or (ii) if $C_m<0$, all right going modes have $M=-i$ and all left going ones have $+i$. Along the mirror symmetric line in SBZ, the Dirac points are symmetric about the origin. A Dirac point can be located either at a TRS momentum or a generic point, and we denote it by $D_0$ and $D$, respectively.

The pairing amplitude at $\bk$ is defined as
\bea
\delta(\bk)=\langle{\Omega}|\hat{\Delta}\psi(\bk)\hat{T}\psi(\bk)\hat{T}^{-1}|\Omega\rangle,
\eea
where $\hat\Delta\equiv\tilde\Delta_{\alpha\beta}c^\dag_\alpha(\bk)c^\dag_\beta(-\bk)+h.c.$ is the pairing operator, $\psi(\bk)$ is the eigenstate annihilation operator in the \emph{normal state} at $\bk$, and $|\Omega\rangle$ is the Fermi liquid ground state (non-superconducting). This amplitude is well defined for any non-degenerate $\bk$-point. Using TRS\cite{Qi2010}, it can be proved that $\delta(\bk)=\delta(-\bk)\in{real}$. To the lowest order of pairing strength, the Bougliubov excitation at $\bk$ has energy $\sqrt{(\epsilon(\bk)-\mu)^2+\delta(\bk)^2}$. This means $\delta(\bk)$ cannot change sign on any connected FS in a gapped superconductor. At the same time, if there are multiple disjoint FS's, then two FS's may have opposite signs, unless they are related to each other by TRS. Sign changes between disjoint FS's usually implies unconventional pairing mechanisms. The 1111-family of iron-based superconductors belongs to this special class. For simplicity, let us for now assume that all FS's have the same sign of pairing amplitude, and the generalization to the generic case is made in Sec.\ref{sec:total_number}.

\subsection{Surface Hamiltonians in the normal state}
We first write down the effective theory ($k\cdot{p}$-model) for each $D_0$ and $D$ without superconductivity. For $D_0$, the little group is generated by $M$ and $T$, represented by $i\sigma_y$ and $K(i\sigma_y)$. So the effective theory reads:
\bea
h_0(\bq)=-\mu\sigma_0+v_1\sigma_yq_x+(v_2\sigma_x+v_3\sigma_z)q_x+O(|q|^2).
\eea
At $D$, the little group is generated by only $M=i\sigma_y$, so
\bea
h(\bq)=(v_0q_x-\mu)\sigma_0+v_1\sigma_yq_x+(v_2\sigma_x+v_3\sigma_z)q_x+O(|q|^2).
\eea

\subsection{Bulk superconductivity for a Dirac cone centered at TRIM}

Next we consider a homogeneous onsite BCS pairing. Since $D_0$ is a TRS momentum, the Cooper pairing is intra-pocket. The BdG Hamiltonian in general looks
\bea\label{eq:BdG_H0}
H_0(\bq)=\left(\begin{matrix}
h_0(\bq) & \tilde\Delta\\
\tilde\Delta^\dag & -h^T_0(-\bq)
\end{matrix}\right).
\eea
The form of $\tilde\Delta$ is determined by the symmetry constraints, namely,
\bea
(i\sigma_y)^T\tilde\Delta(i\sigma_y)&=&\tilde\Delta\textrm{ from mirror symmetry and}\\
\nonumber
(i\sigma_y)^T\tilde\Delta^\ast(i\sigma_y)&=&\tilde\Delta\textrm{ from TRS}.
\eea
These constraints require
\bea
\tilde\Delta=\Delta_2\sigma_0+i\Delta_1\sigma_y,
\eea
where $\Delta_{1,2}\in{Real}$. Diagonalizing the homogeneous Hamiltonian exactly, we obtain
\bea\label{eq:h0_dispersion}
E(\bq)=\pm\sqrt{v_1^2q_x^2+(v^2_2+v_3^2)q_y^2+\Delta_1^2+\Delta_2^2+\mu^2\pm2\sqrt{q_x^2v_1^2\mu^2+q_y^2(v_2^2+v_3^2)(\Delta_2^2+\mu^2)}}.
\eea
It is straightforward to show that the dispersion in Eq.(\ref{eq:h0_dispersion}) is always gapped as far as $\Delta_1^2+\Delta_2^2\neq0$. This means the surface state around $D_0$ is always gapped, against any parameter change. On this gapped FS, we can calculate the sign of the pairing amplitude: $\textrm{sign}(\delta(\bq))=\textrm{sign}(\Delta_1)$.

\subsection{Bulk superconductivity for two Dirac cones centered at $\pm\bD$}

For a Dirac cone around $D$, there must be another around $-D$ due to TRS. The gauge at $D$ is chosen such that
\bea
\hat{M}f_{+,\tau}(\bq)\hat{M}^{-1}=(i\sigma_y)_{\tau\tau'}f_{+,\tau'}(M\bq),
\eea
where $f_{+,\tau}(\bq)$ is the electron annihilation operator at $\bD+\bq$. We can thus fix the gauge at $-\bD-\bq$ by TRS symmetry, i.e., $f_{-,\tau}(\bq)\equiv\hat{T}f_{+,\tau}(-\bq)\hat{T}^{-1}$. Since $[\hat{M},\hat{T}]=0$, at $-\bD$ ($\bq=0$), we also have the representation of the mirror symmetry as $M=i\sigma_y$. In this gauge, the $k\cdot{p}$-model for the cone centered at $-\bD$ is
\bea
h_-(\bq)=-(v_0q_x-\mu)\sigma_0+v_1q_x\sigma_y-(v_2\sigma_x+v_3\sigma_z)q_y.
\eea
We have assumed that Cooper pairs have zero momentum, so the pairing is only between $f_+(\bq)$ and $f_-(-\bq)$, leading to the following BdG Hamiltonian:
\bea\label{eq:BdG_H}
H(\bq)=\left(\begin{matrix}
h_+(\bq) & 0 & 0 & \tilde\Delta\\
0 & h_-(\bq) & -\tilde\Delta^T & 0\\
0 & -\tilde\Delta^\ast & -h^T_+(-\bq) & 0\\
\tilde\Delta^\dag & 0 & 0 & -h^T_-(-\bq)
\end{matrix}\right).
\eea
Eq.(\ref{eq:BdG_H}) is in a block-diagonalized form, and the pairing matrix $\tilde\Delta$ is constrained by
\bea
(i\sigma_y)^T\tilde\Delta(i\sigma_y)&=&\tilde\Delta,\\
\nonumber
\tilde\Delta^\dag&=&\tilde\Delta.
\eea
The only possible form of $\tilde\Delta$ is then $\tilde\Delta=\Delta_1\sigma_0+\Delta_2\sigma_y$, where $\Delta_{1,2}\in{real}$. Eq.(\ref{eq:BdG_H}) can also be exactly diagonalized, giving the dispersion:
\bea\nonumber
E^2(\bq)=v_1^2q_x^2+(v_2^2+v_3^2)q_y^2+(\mu\pm{}v_0q_x)^2+\Delta_1^2+\Delta_2^2
\pm2\sqrt{[\pm{}q_xv_1(\pm{}v_0q_x-\mu)+\Delta_1\Delta_2]^2+q_y^2(v_2^2+v_3^2)[(v_0q_x\pm\mu)^2+\Delta_2^2]}.
\eea
Tedious algebra shows that when $|\Delta_1|>|\Delta_2|\ge0$, the spectrum is fully gapped and when $|\Delta_2|>|\Delta_1|\ge0$, we have four band crossings at zero energy, namely: 
\bea\nonumber
(q_x,q_y)=(\pm\frac{\Delta_1\mu}{v_0\Delta_1-v_1\Delta_2},\pm\sqrt{\frac{(\Delta_2^2-\Delta_1^2)[(v_0\Delta_1-v_1\Delta_2)^2+\mu^2v_1^2]}{|v_0\Delta_1-v_1\Delta_2|(v_2^2+v_3^2)}}).
\eea
This surface superconducting phase has four nodes in four quadrants respectively, and two nodes related by mirror symmetry have opposite winding numbers of $d$-vectors leaving the total winding number zero. But this is not our current interest which is fully gapped surface superconductivity. Therefore, we take $|\Delta_1|>|\Delta_2|$, and the dispersion is always gapped against any change in parameters. We can also calculate the sign of the pairing amplitude, obtaining $\textrm{sign}(\delta(\bq))=\textrm{sign}(\Delta_1)$.

\subsection{Superconducting vortex bound states for a Dirac cone centered at TRIM}

We are now ready to consider a vortex in the Cooper pairing. In principle, a vortex in real space will induce Cooper pairing between Dirac cones that are not opposite to each other in the momentum space, but we will for now ignore this effect, which means that $r_0\delta{D}\gg1$, where $r_0$ is the size of the vortex and $\delta{D}$ is the distance in $k$-space between nearest Dirac cones. After we calculate, under this working assumption, the number of symmetry protected MZM's, we then can argue that so far as the bulk gap is open, this symmetry protected quantum number remains the same under any perturbation that we ignore during the calculation, although the explicit wavefunctions of the MZM's generically vary.

First we study the surface states around $\bD_0$. The vortex is introduced by replacing $\Delta_{1,2}$ in Eq.(\ref{eq:BdG_H0}) by $\Delta_{1,2}(r)e^{-i\theta}$, where $\Delta_{1,2}(0)=0$ and $\Delta(\infty)=\Delta_{1,2}>0$.
Since $H_0$ is gapped for any parameter set, we can always adiabatically tune the parameters to $v_1=-v_2\equiv{v}$, $v_0=v_3=\mu=0$ and $\Delta_2=0$. The number of MZM's for this set of parameters is the same as the number of MZM's for the original parameters, since the surface remains fully gapped during the change. Solving the Schrodinger equation
\bea
\left(\begin{matrix}
0 & -ve^{-i\theta}(\partial_r-\frac{i\partial_\theta}{r}) & 0 & \Delta_1(r)e^{-i\theta}\\
ve^{i\theta}(\partial_r+\frac{i\partial_\theta}{r}) & 0 & -\Delta_1(r)e^{-i\theta} & 0\\
0 & -\Delta_1(r)e^{i\theta} & 0 & ve^{i\theta}(\partial_r+\frac{i\partial_\theta}{r})\\
\Delta_1(r)e^{i\theta} & 0 & -ve^{-i\theta}(\partial_r-\frac{i\partial_\theta}{r}) & 0
\end{matrix}\right)
\left(\begin{matrix}
\psi_1\\
\psi_2\\
\psi_1^\ast\\
\psi_2^\ast
\end{matrix}\right)=0,
\eea
we obtain one MZM mode
\bea
\gamma&=&i\int{d^2\br}[f^\dag_\dw(\br)-f_\dw(\br)]e^{-\int_0^r|\frac{\Delta_1(r')}{v}|dr'},\;\textrm{if}\;v\Delta_1>0;\\
\gamma&=&\int{d^2\br}[f^\dag_\dw(\br)+f_\dw(\br)]e^{-\int_0^r|\frac{\Delta_1(r')}{v}|dr'},\;\textrm{if}\;v\Delta_1<0.
\eea
Using $T=i\sigma_yK$ and $M=i\sigma_y$, it is easy to verify that 
\bea\label{eq:TRIM}
\hat{M}_T\gamma\hat{M}^{-1}_T=\textrm{sign}(v_1\Delta_1)\gamma.
\eea

\subsection{Superconducting vortex bound states for Dirac cones centered at $\pm\bD$}

For the Cooper pairing between the surface electrons around $\bD$ and $-\bD$, the Schrodinger equation can be block diagonalized into two equations, namely
\bea
\left(\begin{matrix}
0 & -ve^{-i\theta}(\partial_r-\frac{i\partial_\theta}{r}) & \Delta_1(r)e^{-i\theta} & 0\\
ve^{i\theta}(\partial_r+\frac{i\partial_\theta}{r}) & 0 & 0 & \Delta_1(r)e^{-i\theta}\\
\Delta_1(r)e^{i\theta} & 0 & 0 & ve^{i\theta}(\partial_r-\frac{i\partial_\theta}{r})\\
0 & \Delta_1(r)e^{i\theta} & -ve^{-i\theta}(\partial_r-\frac{i\partial_\theta}{r}) & 0
\end{matrix}\right)
\left(\begin{matrix}
\psi_1\\
\psi_2\\
\psi_3\\
\psi_4
\end{matrix}\right)=0,
\eea
and
\bea
\left(\begin{matrix}
0 & -ve^{i\theta}(\partial_r+\frac{i\partial_\theta}{r}) & -\Delta_1(r)e^{-i\theta} & 0\\
ve^{-i\theta}(\partial_r-\frac{i\partial_\theta}{r}) & 0 & 0 & -\Delta_1(r)e^{-i\theta}\\
-\Delta_1(r)e^{i\theta} & 0 & 0 & ve^{i\theta}(\partial_r+\frac{i\partial_\theta}{r})\\
0 & -\Delta_1(r)e^{i\theta} & -ve^{-i\theta}(\partial_r-\frac{i\partial_\theta}{r}) & 0
\end{matrix}\right)
\left(\begin{matrix}
\psi_3^\ast\\
\psi_4^\ast\\
\psi_1^\ast\\
\phi_2^\ast
\end{matrix}\right)=0,
\eea
where we have used that MZM must be invariant under PHS. Also, since $M_T$ is a symmetry, the combination $P\times{M}_T$ is also a symmetry that anti-commutes with the Hamiltonian. A MZM must also be an eigenstate of $P\times{M}_T$ with eigenvalue being either $+1$ or $-1$. This places constraints on $\psi_{1,2,3,4}$, namely, $\psi_1=\pm\psi_4$ and $\psi_2=\mp\psi_3$. Solving the equations analytically, we have two MZM solutions:
\bea
\gamma_1&=&\int{d}r^2[f^\dag_{+,\dw}(\br)-f_{-,\up}(\br)+h.c.]e^{-\int_0^r|\frac{\Delta_1(r')}{v}|dr'},\\
\nonumber
\gamma_2&=&i\int{d}r^2[f^\dag_{+,\dw}(\br)-f_{-,\up}(\br)-h.c.]e^{-\int_0^r|\frac{\Delta_1(r')}{v}|dr'},
\eea
if $\textrm{sign}(v\Delta_1)>0$ and
\bea
\gamma_1&=&\int{d}r^2[f^\dag_{+,\dw}(\br)+f_{-,\up}(\br)+h.c.]e^{-\int_0^r|\frac{\Delta_1(r')}{v}|dr'},\\
\nonumber
\gamma_2&=&i\int{d}r^2[f^\dag_{+,\dw}(\br)+f_{-,\up}(\br)-h.c.]e^{-\int_0^r|\frac{\Delta_1(r')}{v}|dr'},
\eea
if $\textrm{sign}(v\Delta_1)<0$. It is straightforward to verify:
\bea\label{eq:nonTRIM}
\hat{M}_T\gamma_{1,2}\hat{M}^{-1}_T=\textrm{sign}(v\Delta_1)\gamma_{1,2}.
\eea

\subsection{Total number of MZM's}
\label{sec:total_number}
From the analysis, we see that there is one MZM from a Dirac cone centered at a TRIM, and two MZM's from a pair of Dirac cones centered at $\pm\bD$. Due to our assumption that there is no sign change between pairing amplitudes on different FS's, $\Delta_1$ has the same sign for all Dirac cones, therefore, every MZM transforms under $M_T$ as
\bea\label{eq:temp1}
\hat{M}_T\gamma_{i=1,...,|C_m|}\hat{M}^{-1}_T=\textrm{sign}(C_m\Delta_1).
\eea
In the basis of $\gamma_i$'s, the symmetry operation $M_T$ is simply $\mathcal{M}=\textrm{sign}(C_m\Delta_1)I_{|C_m|\times|C_m|}$, so $|\tr(\mathcal{M})|=|C_m|$. Therefore, there are exactly $|C_m|$ MZM's protected by $M_T$.

If there are two mirror symmetric planes in the 3D BZ, then there are two mirror Chern numbers denoted by $C_{m0}$ and $C_{m\pi}$. There are two mirror symmetric lines in the SBZ, along which there are $|C_{m0}|$ and $|C_{m\pi}|$ Dirac points, respectively. An analysis same as the above and the result in Eq.(\ref{eq:temp1}) apply to Dirac cones on both lines. The total number of protected MZM's is hence
\bea
N_{MZM}=|C_{m0}+C_{m\pi}|.
\eea

It is easy to generalize to cases where the different surface Fermi pockets have different signs of pairing amplitude. There are in total $|C_m|$ Fermi pockets on the surface, and the pairing signs are symmetric about the origin due to TRS. The total number of MZM is simply the sum of the signs of all the pockets, namely
\bea
N_{MZM}=|\sum_{m}\textrm{sign}(\Delta_m)|,
\eea
where $\textrm{sign}(\Delta_m)$ is the pairing sign of the $m$-th Fermi pocket.

\section{The two `semi'-Majorana modes in pure samples}

Out of the four MZM's in Eq.(10) of the main text, two can be gapped out by the following perturbation
\bea\label{eq:perturbation}
\delta{H}&=&i\lambda(\gamma_1\gamma_2+\gamma_2\gamma_3+\gamma_3\gamma_4+\gamma_4\gamma_1)\\
\nonumber
&=&i\lambda(\gamma_1-\gamma_3)(\gamma_2-\gamma_4).
\eea
From Eq.(10) in the main text, we can see that $\gamma_1-\gamma_3$ is composed of electrons from Dirac cones centered at $\bD_{1,3}$, and $\gamma_2-\gamma_4$ is composed of electrons from Dirac cones centered at $\bD_{2,4}$. Since $\bD_{1,3}\pm\bD_{2,4}\sim(\pi,\pi)$, all terms in Eq.(\ref{eq:perturbation}) carry large momentum scattering or large momentum pairing. Large momentum scattering is suppressed if the impurity potential is smooth; large momentum pairing is suppressed if the vortex size is much larger than the lattice constant. When both effects are suppressed, the coefficient $\lambda$ in Eq.(\ref{eq:perturbation}) must be very small. Since Eq.(\ref{eq:perturbation}) is the only allowed hybridization term, there are four, instead of two MZM's.

\end{document}